\documentstyle[11pt,newpasp,twoside,epsf]{article}
\begin{document}
\title{Tidal Streams and Low Mass Companions of M31}
 \author{Robert Braun}
\affil{ASTRON, P.O. Box 2, 7990 AA Dwingeloo, The Netherlands}
\author{David Thilker}
\affil{Department of Physics and Astronomy,
Johns Hopkins University, 
3400 N. Charles St., 
Baltimore MD 21218-2695, U.S.A.}

\begin{abstract}
We have imaged the extended HI environment of M31 with an unprecedented
combination of high resolution and sensitivity. We detect a number of
distinct High Velocity Cloud components associated with M31. A sub-set
of the features within 30~kpc appear to be tidal in origin. A
filamentary ``halo'' component is concentrated at the M31 systemic
velocity and appears to extend into a ``bridge'' connecting M31 and
M33. This may represent condensation in coronal gas. A population
of discrete clouds is detected out to radii of about 150~kpc. Discrete
cloud line-widths are correlated with HI mass and are consistent with a
100:1 ratio of dark to HI mass. These may be the gaseous counterparts
of low-mass dark-matter satellites. The combined distribution of M31's HVC
components can be characterized by a spatial Gaussian of 55~kpc
dispersion and yields an N$_{HI}$ distribution function which agrees
well with that of low red-shift QSOs.
\end{abstract}

\section{Introduction}

The high velocity cloud (HVC) phenomenon has been under study for some
40 years, since the first detections of $\lambda$21~cm emission from
atomic hydrogen at velocities far removed from those allowed by
rotation in the Galaxy disk (Muller, Oort \& Raimond 1963). Several
explanations have been put forth for their interpretation, including a
galactic fountain, infall of circum-galactic gas, tidal debris from
mergers and sub-galactic-mass companions. These last two processes in
particular have a direct bearing on this meeting. The suggestion has
been made that at least one component of the Galactic HVCs, the
so-called CHVCs (Braun \& Burton 1999, Blitz et al. 1999) might be the
gaseous counterpart of low-mass dark-matter satellites. A definitive
explanation for the Galactic HVCs is hampered by the extreme difficulty
of obtaining direct distance estimates. Although this difficulty can,
in principle, be circumvented by studying external galaxies, it has
proven extremely difficult to achieve the necessary combination of high
spatial resolution, low mass sensitivity and large field-of-view in
practise (Braun \& Burton 2001).  HI studies of nearby galaxies have
often revealed features that could be termed HVCs (eg. in NGC~628 by
Kamphuis \& Briggs 1992), but typically only one or a few of such
features have been detected near each host at masses as low as about
10$^7$~M$_\odot$. Only in the last few years has sufficient sensitivity
and resolution been obtained to begin revealing diffuse systems of
peculiar velocity HI, such as the slowly rotating ``halo'' component of
NGC~2403 (Fraternali et al. 2002).

We have undertaken a program of three seperate surveys (defined below)
to determine the HI content in the environment of our nearest spiral
neighbour, M31, with high sensitivity and over a very wide range of
spatial scales. This has allowed us to detect several distinct classes
of HVC components associated with M31. Indications are seen for most or
all of the hypothesized HVC explanations listed above. While the origin
of all components is not yet firmly established, the relative importance of
the various processes is at last beginning to be constrained.

\begin{figure}
\plotone{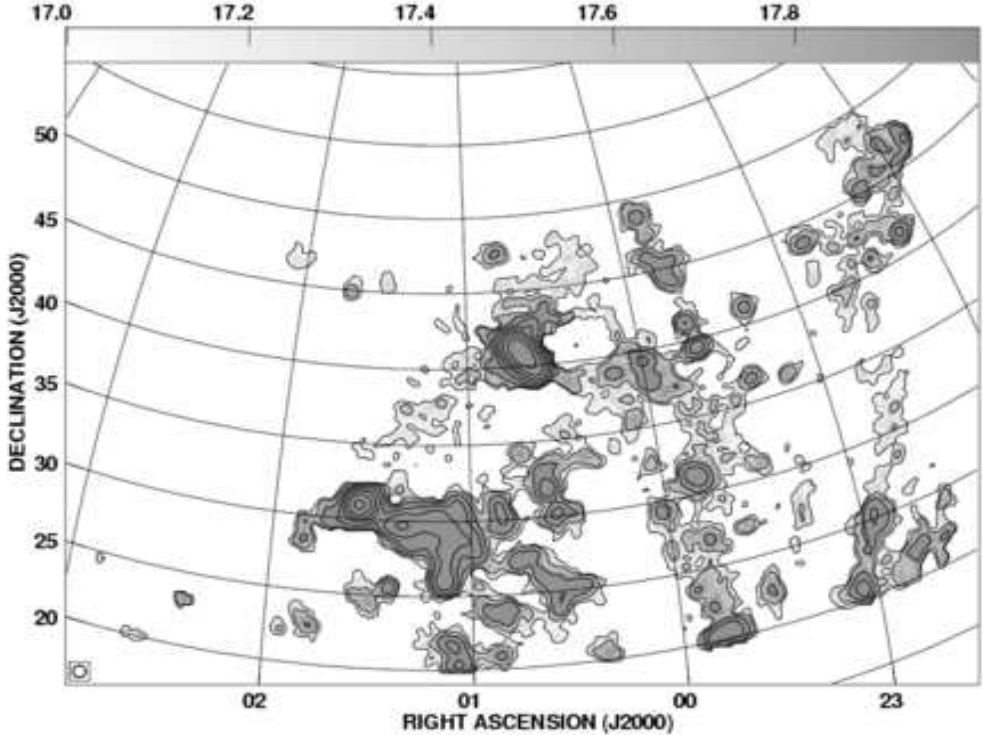}
\caption{Integrated high velocity HI within the WSRT wide-field survey
region. The grey-scale varies between
log(N$_{HI}$)~=~17~--~18. Contours are drawn at log(N$_{HI}$)~=~17,
17.5, 18, $\dots$ 20.5. }
\end{figure}

%\begin{figure}  
%\plotone{m31smpzb.mpr.eps}
%\caption{Peak intensity of HI emission in the WSRT mosaic of M31. 
%The
%  displayed brightness temperatures vary between about 2 and 145 Kelvin
%  and are displayed with a square root transfer function. 
% }
%\end{figure}

\section{Three HI Surveys of the M31 Environment}

\subsection{The WSRT wide-field Survey}

The WSRT array was used to acquire 380 hours of observing in the period
Aug. 2002 to Oct. 2002 to complete a drift-scan survey centered
approximately on the position of M31 and Nyquist-sampled over
60$\times$30\deg in R.A.$\times$Dec. The primary data consists of
auto-correlation spectra with an effective angular resolution of
49$\arcmin$ FWHM, although cross-correlation data were also acquired
simultaneously. An {\sc rms} sensitivity of about 18~mJy/Beam at a
velocity resolution of 17~km/s was realized between
$-1000~<~$V$_{Hel}~<~+6500$~km/s. Further details of this survey
together with an analysis of the detected background galaxies can be
found in Braun, Thilker \& Walterbos (2003). The analysis of the Local
Group features is presented in Braun \& Thilker (2003). At the distance
of M31, the angular resolution and field-size correspond to about
11~kpc over a region of 800$\times$400~kpc, while the RMS column
density sensitivity was 4$\times$10$^{16}$cm$^{-2}$. This can be
compared to the recently completed HIPASS survey (Barnes et al. 2001)
which reaches a similar point source flux sensitivity of 14~mJy/Beam at
a velocity resolution of 18~km/s in a 15\farcm5 beam, but an order of
magnitude poorer column density sensitivity of about
4$\times$10$^{17}$cm$^{-2}$.

An overview of detected emission features which were distinct from the
intermediate velocity wings of Galactic emission at negative LSR
velocities is shown in Figure~1. We detect high velocity HI emission at
a column density in excess of our 2$\sigma$ limit of about
10$^{17}$cm$^{-2}$ from fully 29\% of the area of our 1800 deg$^2$
survey field. The detected emission is due to a mix of discrete and
diffuse components. The brightest discrete features visible in this
image are M31 ($\alpha,\delta$)=(00:43,+41:16)), M33 (01:34,+30:40) and
Wright's HVC (01:15,+29:00). Some 95 additional discrete HVCs are
detected, of which more than 80\% are isolated from any diffuse HI
emission at a column density of 1.5$\times$10$^{17}$cm$^{-2}$. 

Diffuse HI emission is also detected in the form of: {\bf 1)} a
``western filament'' extending north-north-west from
($\alpha,\delta$)=(23:00,+20) to (22:30,+35), {\bf 2)} an ``eastern
loop'' extending to the north from (23:40,+20) to (23:40,+42) and
arcing back down to (01:00,+20) and {\bf 3)} a diffuse bridge of
emission that extends between M31 and M33 and continues to the
north-west of M31 to ($\alpha,\delta$) = (00:20,+48). The line-of-sight
velocity varies continuously along each of these features, with the
M31/M33 ``bridge'' connecting the systemic velocities of these two
galaxies.

\begin{figure}
\plotone{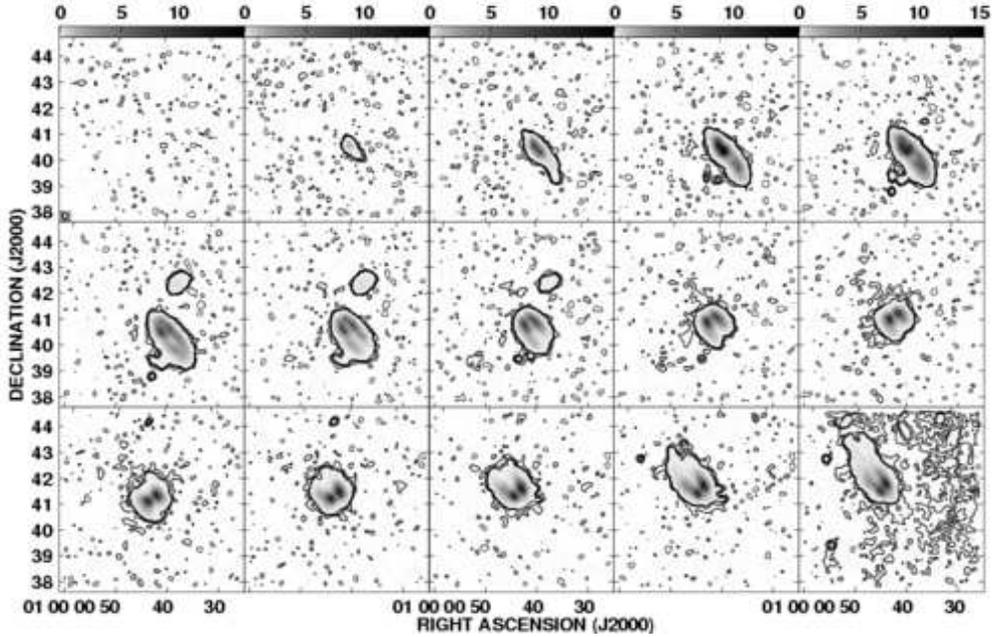}
\caption{Channel maps of HI emission in the GBT survey. Contours are 
drawn at $-$2, 2, 4, and 6$\sigma$.  
%Note the population of
%discrete high-velocity clouds, plus a filamentary halo distribution of
%gas near M31's systemic velocity ($-$300 km/s).  
Velocities are: $-$689 (top left),
$-$656, \dots $-$181 km/s (bottom right).}
\end{figure}

\subsection{The GBT Survey}

The recently completed GBT was used for about 70 hours observing in the
period July 2002 to Sept. 2002 to complete six Nyquist-sampled
on-the-fly coverages of a 7$\times$7\deg field centered on M31 (as
well as a 5$\times$5\deg field centered on M33, which we will discuss
elsewhere). An {\sc rms} sensitivity of about 6.6~mJy/Beam was realized
at a velocity resolution of 18~km/s (although the velocity resolution
was 1~km/s). The angular resolution and field-size correspond to
about 2~kpc over a region of 95$\times$95~kpc, while the RMS column
density sensitivity was about 1.5$\times$10$^{17}$cm$^{-2}$ (at 18~km/s). 
Further details of this survey and it's analysis can be found in
Thilker et al. (2003).

An overview of the GBT survey data is given in Figure~2, as a series of
channel maps after a mild spatial and velocity smoothing (to 3~kpc,
36~km/s). A population of at least 20 faint and compact discrete HI
features is apparent at heliocentric velocities between about $-$520
and $-$180~km/s. Galactic foreground confusion limits detections to
velocities more negative than about $-$180~km/s. In addition, a diffuse
filamentary component can be seen centered spatially on M31 near the
systemic velocity of $-$300~km/s.

\subsection{The WSRT Mosaic}

The WSRT array was used to acquire some 350 hours of observing in the
period Aug. 2001 to Jan. 2002 distributed over 163 pointing centers on
a 6$\times$3\deg Nyquist-sampled grid aligned with the M31 major
axis. Joint deconvolution of this synthesis data, together with total
power data, required some 3 months processing on a 4-CPU machine that
was kindly made available to us for this purpose by the PuMa team of
the University of Amsterdam. An {\sc rms} sensitivity of about
1.4~mJy/Beam was realized at a velocity resolution of 2~km/s. The
maximum angular resolution of 15\arcsec corresponds to about 50~pc
resolution over a region of 80$\times$40~kpc. The column density
sensitivity (at 20~km/s) is 1.0, 3.5, 11 and
24$\times$10$^{18}$cm$^{-2}$ at angular resolutions of 120, 60, 30 and
20\arcsec. A more extensive description of the survey and it's
scientific goals can be found in Braun et al. (2002, 2003). 

The combined data is presented in Braun et al. (2002), where the peak
brightness temperature is shown at the full spatial and velocity
resolution. In addition to a very detailed view of the M31 disk (the
largest ratio of FOV to beam-size yet acquired in HI for any galaxy),
the compact cores are detected of many HVC features which are 10's of
kpc from the nearest normal disk emission.

\section{Results}

\subsection{Foreground Features: The Magallenic Stream}

As can be seen in Fig.~1, high velocity HI features are almost
ubiquitous from our vantage point in the Galaxy, once sufficient column
density sensitivity is employed. Fortunately, the line-of-sight
velocity can be used to distinguish different components from one
another. Of the three diffuse features noted previously in the
wide-field survey field-of-view, the first two (the ``western
filament'' and ``eastern loop'') constitute a previously undetected
extension of the Magellanic Stream (MS).

\begin{figure}
\plotone{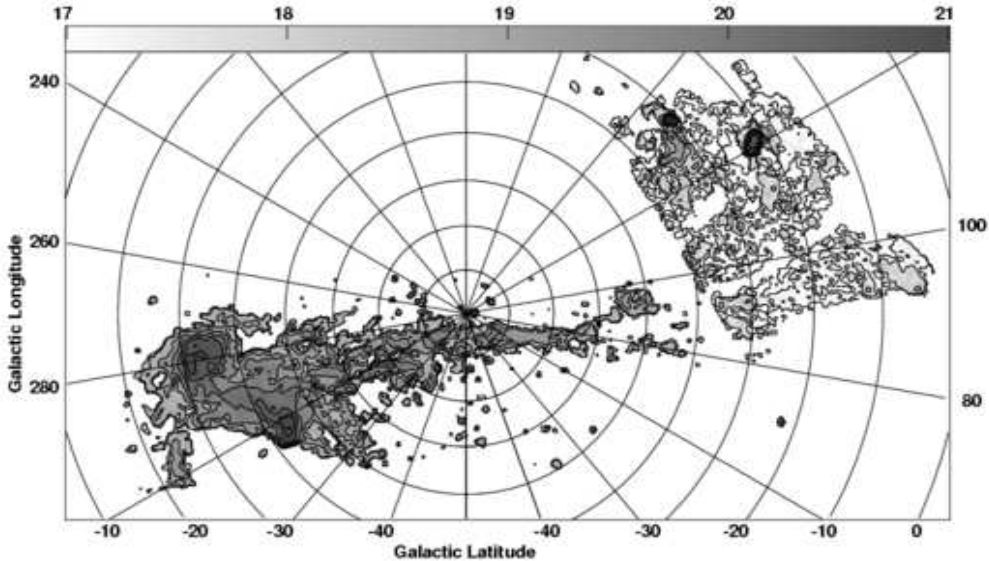}
\caption{Integrated HI emission of the Magellanic Clouds and
Stream. The HIPASS data of Putman et al. (2003) is combined with the
wide-field WSRT survey data ($90<l<160, -40<b<-5$). The
grey-scale range is log(N$_{HI}$)~=~17~--~21, for N$_{HI}$ in
units of cm$^{-2}$. Contours are drawn at log(N$_{HI}$)~=~17.5, 18.5,
19.0, 19.5, 20.0 \dots 21.5. }
\end{figure}

The full extent of the Stream can be seen in Fig.~3, where we have
combined the HIPASS data of Putman et al. (2003) with the WSRT
wide-field survey. Note that the HIPASS data is limited by sensitivity
to log(N$_{HI}$)~$>$~18.3 while the WSRT data reach an order of
magnitude deeper for these diffuse structures, to
log(N$_{HI}$)~$>$~17.3. It is important to realize that only a sub-set
of the diffuse features seen in this figure are kinematically
associated with the MS (Braun \& Thilker 2003). This sub-set displays a
smoothly varying, single-valued radial velocity. Additional diffuse and
discrete components are detected which may well have other origins.

Both the approximate spatial distribution and the radial velocity of
the ``western filament'' and ``eastern loop'' extensions of the MS were
correctly predicted in the models of Gardiner \& Noguchi (1996) and
Gardiner (1999). These models seem to require the inclusion of some ram
pressure interaction with the Galactic halo although they are dominated
by gravitational interaction. In these models the orbit of the
Magellanic Clouds with the Galaxy varies between about 150 and 50~kpc
radius. The Clouds are currently near peri-galacticon, while the most
distant trailing portions of the Stream (which we have now detected)
trace the apo-galacticon portion of the orbit near 150~kpc, where the
Clouds where located some 0.9~Gyr ago.

\subsection{The M31 HVC Populations}

Several different components of high velocity HI are detected in our
three surveys which appear to be intimately related to M31. The largest
of these is the diffuse bridge of HI connecting the systemic velocities
of M31 and M33 and continuing to the north-west of M31 in the anti-M33
direction. Superposed on this distribution is a population of very
faint, discrete HI clouds with a very broad range of radial velocities
(at least $-$520 to $-$180~km/s) that is only detected within a radius
of about 15\deg (corresponding to 200~kpc) of M31. This can be compared
with the normal disk rotation of M31 which extends from about $-$600 to
$-$20~km/s. The most extreme negative velocities of previously
cataloged (C)HVCs (De Heij et al. 2002, Putman et al. 2002) are also in
this general direction, but they do not exceed $-$465~km/s. The M31
components are shown in left-hand panel of Fig.~4 as they appear in our
wide-field survey after blanking of all features which have kinematic
continuity with other, presumed fore-ground, HVC features; most notably
the Magellanic Stream and Wright's Cloud.

\begin{figure}  
\plottwo{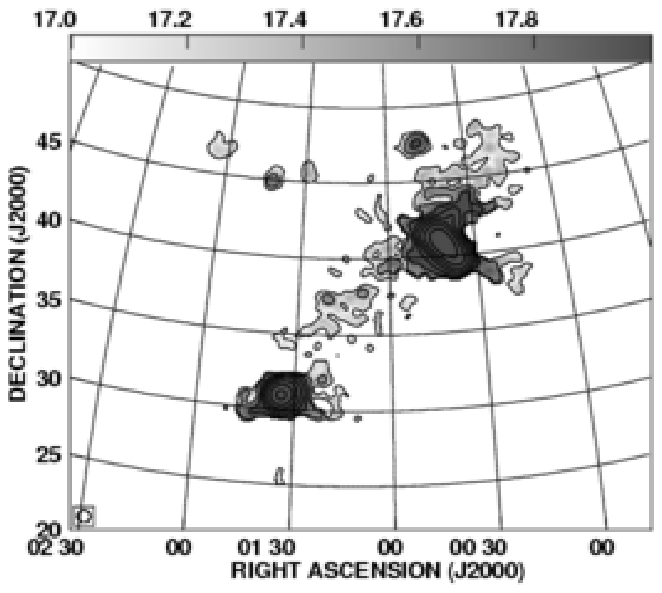}{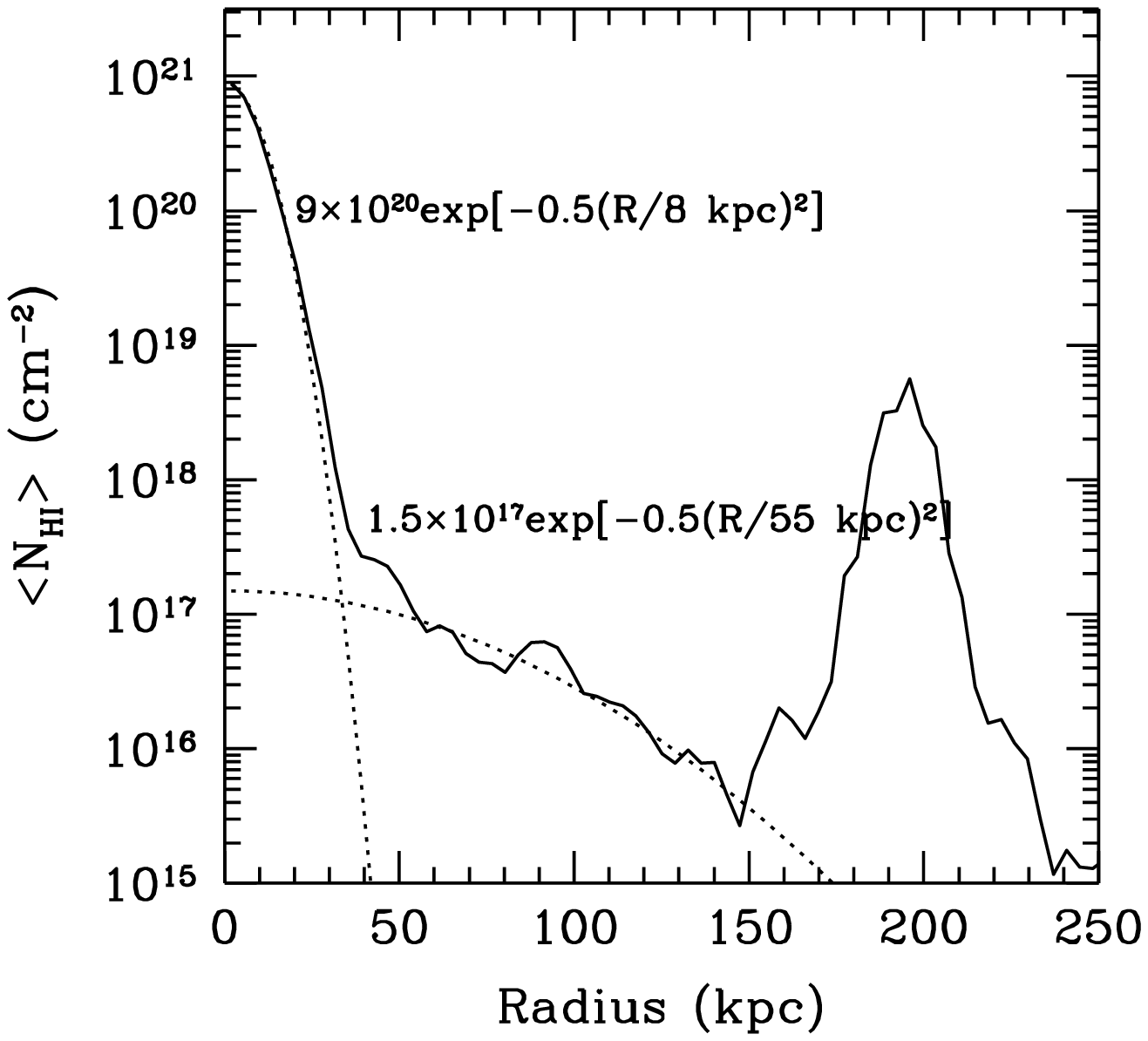}
\caption{{\bf (Left)} Integrated HI emission from the WSRT
 wide-field features kinematically associated with M31 and M33.
 The grey-scale varies between log(N$_{HI}$)~=~17~--~18, for N$_{HI}$
 in units of cm$^{-2}$. Contours are drawn at log(N$_{HI}$)~=~17, 17.5,
 18, $\dots$ 20.5.{\bf (Right)} Azimuthally averaged HI column density
 in the vicinity of M31 as function of radius . The secondary peak near
 200~kpc is due to M33. Two Gaussians are overlaid:
 a central disk component with $\sigma$=8~kpc and peak
 log(N$_{HI}$)~=20.95 as well as an extended circum-galactic component
 of $\sigma$=55~kpc and peak log(N$_{HI}$)~=17.2.}
\end{figure}

The faint HVC poulation near M31 is confirmed in our GBT data of the
central 95$\times$95~kpc. All features in this survey which could be
cleanly separated from the M31 disk emission are shown in Fig.~5
overlaid on a V-band image of the galaxy. The radial velocities of the
discrete HVCs reflect the general pattern of M31 disk rotation, with
the most negative occurring in the south-west and the most positive in
the north-east. A sub-set of the discrete features are organized into
elongated structures with a smoothly varying radial velocity,
suggestive of a tidal origin. The series of features, extending to
about 2\fdg5 (35~kpc) south of the M31 nucleus is the most obvious of
these. This feature is partially coextensive with the ``giant stellar
stream'' of M31 discovered by Ibata et al. (2001). Another feature
seems likely to be related to NGC~205, given it's modest offset in
position and velocity from that galaxy. While this sub-set of features
is strongly suggestive of a tidal origin, the same indications do not
apply to the majority of detected features.

\begin{figure}
\plotone{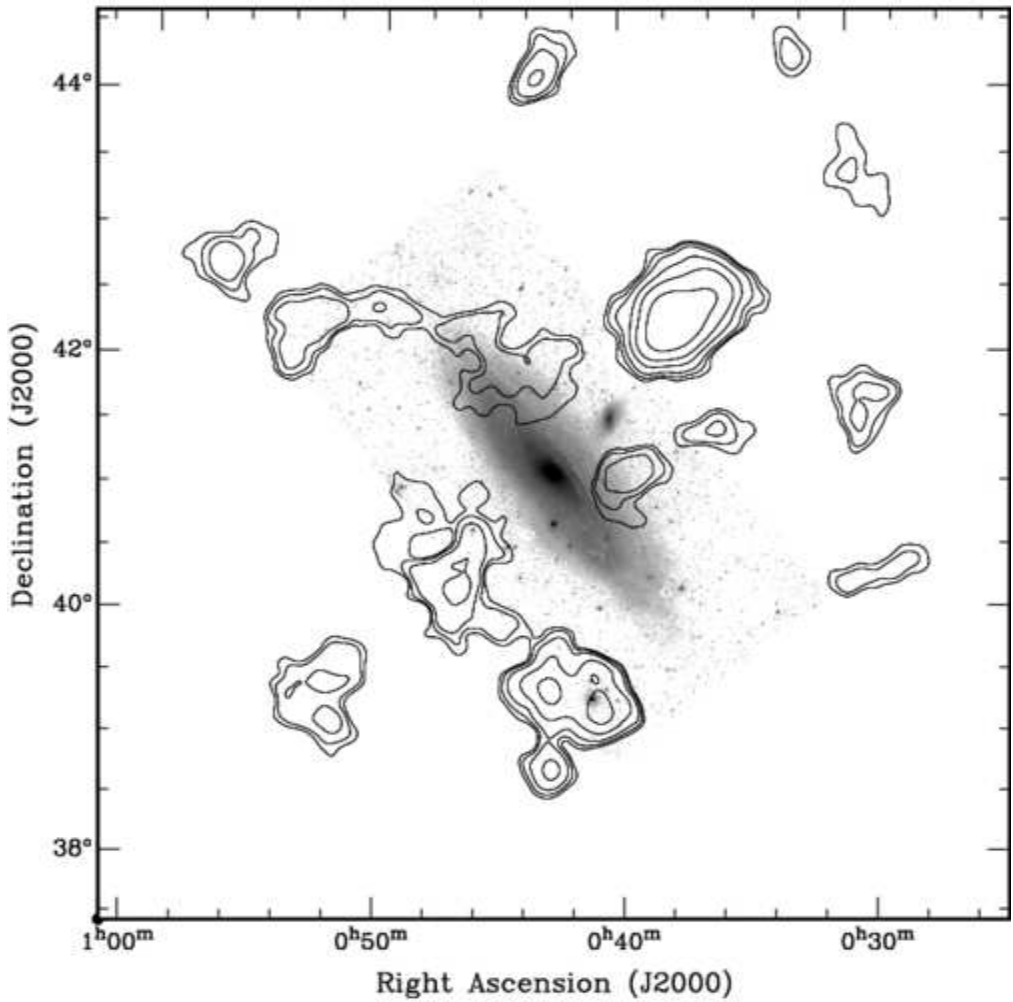}
\caption{Contours of total HI column density for discrete and diffuse
high-velocity HI in the M31 GBT field, after masking out
emission from Andromeda's inclined, rotating disk.  The contours are
drawn at 0.5, 1, 2, 10, and $20\times10^{18}$~cm$^{-2}$, and overlaid
on a V band image of M31. }
\end{figure}

The GBT survey also detects a centrally concentrated filamentary
component near the M31 systemic velocity, reminiscent of the ``bridge''
feature seen on larger scales. We have recently been able to confirm
our faintest discrete HVC detections, together with the brightest
portions of the ``bridge'' with additional pointed GBT observations.
The origin of these diffuse HVC components is not clear, although they
may be a manifestation of cooling in a tenuous inter-galactic medium.
In this view, HVCs condense from, and remain confined by, the bulk of
coronal gas pervading the Local Group. Sembach et al. (2003) present
evidence for an extended and highly ionized Galactic corona or Local
Group medium traced by high-velocity OVI absorption.

The large-scale distribution of the M31 HVC population is illustrated
in the right-hand panel of Fig.~4, where HI column density has been
azimuthally averaged and plotted as a function of projected radius. The
secondary peak in the distribution near 200~kpc is due to M33. Two
Gaussian curves are overlaid on the distribution; a central disk
component with spatial disperion, $\sigma$=8~kpc and peak
log(N$_{HI}$)~=20.95 as well as an extended circum-galactic component
of $\sigma$=55~kpc and peak log(N$_{HI}$)~=17.2. As can be seen in the
left-hand panel of Fig.~4 as well as Fig.~5, the HVC distribution is
actually a combination of discrete and (elliptical) diffuse components,
rather than simply a symmetric Gaussian.

A possible source of high velocity HI near M31 might be the gas
associated with a putative population of low mass dark-matter
halos. Current numerical simulations of the Local Group in a
$\Lambda$CDM cosmology predict a large population of low mass DM halos
surviving to the present epoch (Klypin et al. 1999, Moore et al. 1999)
which dramatically outnumber known dwarf galaxies. If the M31 HVCs were
tracers of substantial dark-matter concentrations, then this should be
directly reflected in their internal linewidths. We plot the observed
HI mass versus FWHM line-width for discrete clouds in the left-hand
panel of Fig.~6.  The line-width distribution has an approximate lower
bound roughly consistent with thermal broadening for gas at 10$^4$ K
(FWHM = 24~km/s). A systematic increase in FWHM line-width is observed
with increasing HI mass, reaching some 70 km/s for the clouds near
10$^6$~M$_\odot$.  Although Davies' HVC is significantly offset from
the remainder of the distribution in the figure, high-resolution
imaging has revealed much higher internal line-widths in that object as
well De Heij et al. (2002). To demonstrate the expected distribution of
line-width with mass, we have over-plotted a curve corresponding to
$V^{2} = 100 G M_{HI} R^{-1}$, where the characteristic discrete cloud
radius, R, has been held constant at 500 pc, based on our WSRT mosaic
detections of some of the cloud cores. This curve, corresponding to a
dark to HI mass ratio of about 100:1, is not intended to represent a
fit to the observed data, but merely provide a basis for
comparison. The hypothesis of a kinematically dominant dark-matter
component appears to be consistent with the observed line-widths of the
discrete M31 HVCs.

Another aspect of the dark-matter mini-halo scenario that can be
checked is the expected number of such objects in the appropriate mass
range in the vicinity of M31. Sternberg et al. (2002) predict $\sim 25$
dark matter mini-halos associated with gravitationally confined HI
within a radius of 40 kpc around M31, based on the simulations of
Klypin et al. (1999) and Moore et al. (1999). This is fully consistent
with the 20 discrete M31 HVC's we have detected.  The Sternberg et
al. calculations also suggest that circum-galactic objects of such low
mass and peak HI column density should be only $\sim10$\%
neutral, implying $M_{HI+HII} \sim 10 M_{HI}$ and $M_{DM} \sim 10
M_{HI+HII}$. 

\begin{figure}
\plottwo{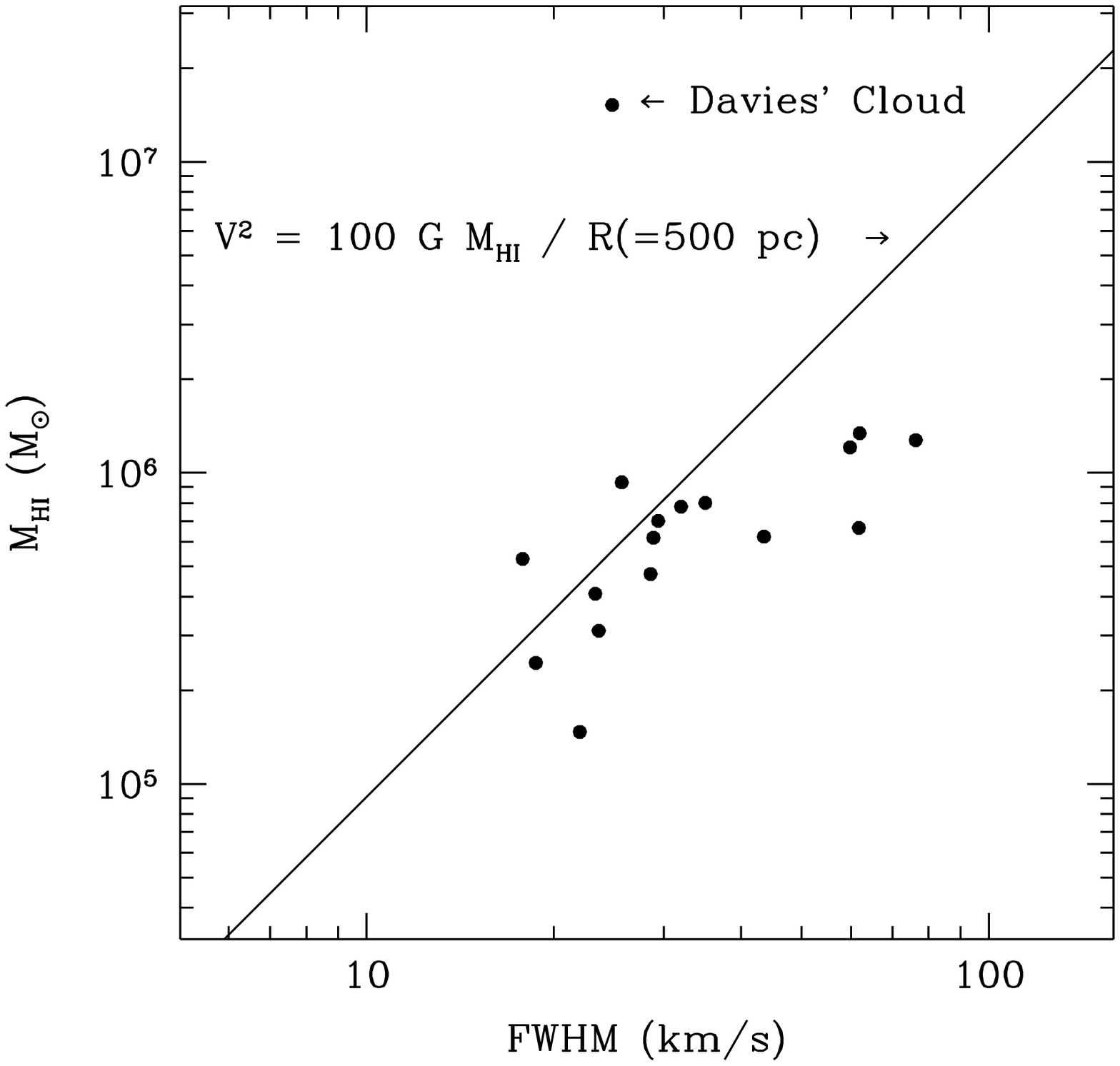}{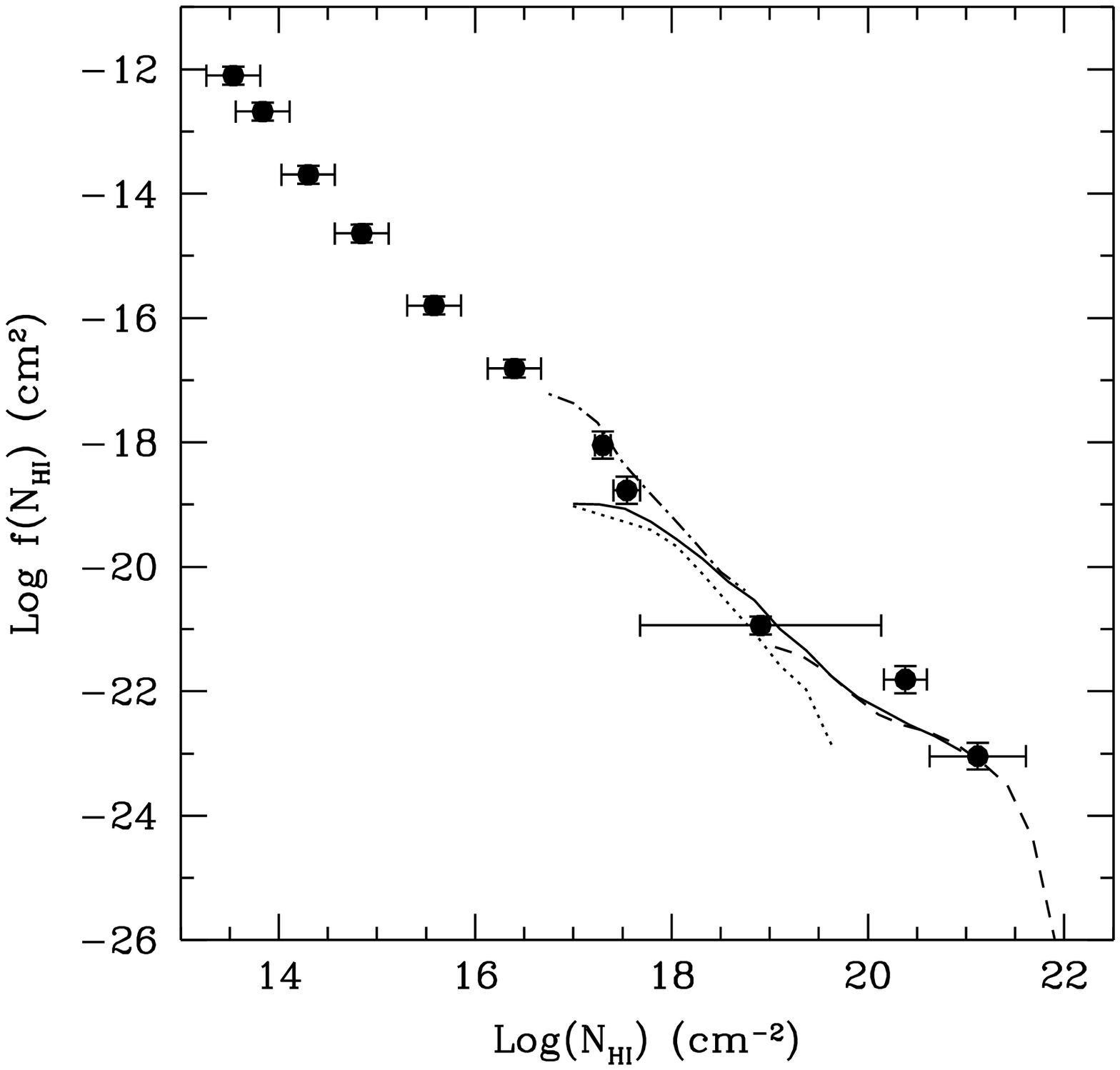}
 \caption{{\bf (Left)} Observed HI mass versus FWHM line-width for
discrete high-velocity clouds in the GBT survey of M31. For comparison,
we plot the expected line-width of rotationally supported objects with
a factor of 100 more mass than detected in HI and a characteristic size
of 0.5 kpc.
%, which follows from our high resolution imaging of several
%cloud cores.  
{\bf (Right)} The distribution function of HI column
density due to M31 and it's HVC population. 
%The data from all three HI
%surveys of M31 are combined in this figure to probe column densities
%over a total range of some five orders of magnitude. The dashed line is
%from the WSRT mosaic (Braun et al. 2003) with 1$\arcmin$ resolution
%over 80$\times$40~kpc, the dotted and solid lines from our GBT survey
% (Thilker et al. 2003) with 9$\arcmin$ resolution over 85$\times$85~kpc
%and the dot-dash line from the wide-field WSRT survey with 48$\arcmin$
%resolution out to 150~kpc radius. 
The filled circles with errorbars are the low red-shift QSO absorption
line data as tabulated by Corbellli \& Bandiera (2002).}
\end{figure}

\subsection{M31 HVCs and QSO Absorption Lines}

Our results for M31 can be placed in a broader context by comparing the
properties of the extended HVC population we detect with those probed
by observations of Ly$\alpha$ absorption toward low red-shift
QSOs. Charlton et al. (2000) have previously suggested making such a
comparison. We have calculated the distribution function of HI column
density by combining the data from all three of our HI surveys of M31
in the right-hand panel of Fig.~6. The dashed line is from the WSRT
mosaic with 1$\arcmin$ resolution over 80$\times$40~kpc, the dotted and
solid lines from our GBT survey with 9$\arcmin$ resolution over
95$\times$95~kpc and the dot-dash line from the wide-field WSRT survey
with 48$\arcmin$ resolution out to about 150~kpc projected radius. The
column density data was normalized with the space density of galaxies
of M31's HI mass (using the HIPASS BGC HIMF of Zwaan et al. 2003).  The
filled circles with errorbars in the figure are the low red-shift QSO
absorption line data as tabulated by Corbellli \& Bandiera (2002).  As
can be seen in the figure, very good agreement is found in both the
shape and normalization of the distribution function. It seems that the
upper five orders of magnitude of the HI distribution function, between
about log(N$_{HI}$)~=~17 and 22, can be accounted for by comparable HVC
populations associated with M$_*$ galaxies.


\begin{references}
\reference Barnes, D.\,G.,  Staveley--Smith, L.,  de Blok, W.\,J.\,G.,
et al. 2001, MNRAS, 322, 486  
\reference Blitz, L., Spergel, D.\,N., Teuben, P.\,J. et al. 1999, ApJ,
514, 818
\reference Braun, R., Burton, W.\,B. 1999, A\&A, 341, 437
\reference Braun, R., Burton, W.\,B. 2001, A\&A, 375, 219
\reference Braun, R., Thilker, D.\,A., Walterbos, R.\,A.\,M., 2003,
A\&A, 406, 829
\reference Braun, R., Thilker, D.\,A. 2003, A\&A, submitted
\reference Braun, R., Thilker, D.\,A., Corbelli, E., Walterbos,
R.\,A.\,M. 2002, \\ 
http://www.astron.nl/newsletter/2002-1/index.html
\reference Braun, R., Thilker, D.\,A., Corbelli, E., Walterbos,
R.\,A.\,M. 2003, A\&A, in prep.
\reference Charlton, J.\,C., Churchill, C.\,W., Rigby, J.\,R. 2000, ApJ
544, 702
\reference Corbelli, E., Bandiera, R. 2002, ApJ, 567, 712
\reference de\,Heij, V., Braun, R., Burton, W.\,B. 2002, A\&A, 391, 67
\reference de\,Heij, V., Braun, R., Burton, W.\,B. 2002, A\&A, 391, 159
\reference Gardiner, L.\,T., Noguchi, M. 1996, MNRAS, 278, 191
\reference Gardiner, L.\,T.,  1999, in ASP Conf. Ser. Vol. 166,
High-Velocity Clouds, eds. B. Gibson \& M. Putman (San Franciscio: ASP), 292
\reference Fraternali, F., van Moorsel, G., Sancisi, R., Oosterloo,
T. 2002, AJ, 123, 3124
\reference Ibata, R., Irwin, M., Lewis, G. et al. Nature, 412, 49 
\reference Kamphuis, J. \& Briggs, F. 1992, A\&A, 253, 335
\reference Klypin, A., Kravtsov, A. V., Valenzuela, O., Prada, F. 1999,
ApJ, 522, 82
\reference Moore, B. et al. 1999, \apj, 524, L19
\reference Muller, C.\,A., Oort, J.\,H., \& Raimond, E. 1963,
C.R.Acad.Sci. Paris, 257, 1661
\reference Putman, M.\,E. de\,Heij, V., Stavely$-$Smith, L., et
al. 2002, AJ, 123, 873 
\reference Putman, M.\,E., Stavely$-$Smith, L., Freeman, K., et al.
2003, ApJ, 586, 170 
\reference Sembach, K.\,R., Wakker, B.\,P., Savage, B.\,D. 2003, ApJS, 146, 165
\reference Sternberg, A., McKee, C.\,F., Wolfire, M.\,G. 2002, ApJS, 143, 419
\reference Thilker, D.\,A., Braun, R., Walterbos, R.\,A.\,M., et
al. 2003, ApJL, in press
\reference Zwaan, M.\,A., Staveley--Smith L.,  Koribalski, B.\,S., et
al. 2003, AJ, 125, 2842

\end{references}
\end{document}